\documentclass[twocolumn,showpacs,preprintnumbers,amsmath,amssymb,prb]{revtex4}% Physical Review B
\usepackage{graphicx}% Include figure files
\usepackage{dcolumn}% Align table columns on decimal point
\usepackage{bm}% bold math

\begin{document}

\title {Superconducting phase diagram of the filled skuterrudite 
PrOs$_4$Sb$_{12}$}
\author{ M.-A. Measson$^1$, D. Braithwaite$^1$, J. Flouquet$^1$, G. Seyfarth$^2$, J.P.~Brison$^2$, 
E.~Lhotel$^2$, C.~ Paulsen$^2$, H.~ Sugawara$^3$, and H.~ Sato$^3$}
\affiliation{
$^{1}$D\'{e}partement de Recherche Fondamentale sur la 
Mati\`{e}re Condens\'{e}e,SPSMS, CEA Grenoble, 38054 Grenoble, France\\ 
$^{2}$Centre de Recherches sur les Tr\`{e}s basses 
temp\'{e}¥ratures, CNRS, 25 avenue des Martyrs, BP166, 38042 Grenoble CEDEX 9, France\\
$^{3}$Department of Physics, Tokyo Metropolitan University, Minami-Ohsawa 1-1, Hashioji, Tokyo 192-0397, Japan}

\begin{abstract}

We present new measurements of the specific heat of the heavy 
fermion superconductor PrOs$_{4}$Sb$_{12}$, on a sample which 
exhibits two sharp distinct anomalies at $T_{c1}= 1.89$K and 
$T_{c2}= 1.72$K. They are used to draw a precise
 magnetic field-temperature superconducting phase diagram of 
PrOs$_{4}$Sb$_{12}$ down to 350~mK.
 We discuss the superconducting phase diagram of PrOs$_{4}$Sb$_{12}$ 
and its possible relation with an 
unconventional superconducting order parameter. We give a detailed 
analysis of  $H_{c2}(T)$, 
which shows paramagnetic limitation (a support for even parity 
pairing) and multiband effects.

\end{abstract}
\maketitle

\section{Introduction}
The first Pr-based heavy fermion (HF) superconductor
PrOs$_{4}$Sb$_{12}$ (T$_{c}\sim 1.85$K) has been recently
discovered \cite{Bauer2002}. Evidence for its heavy fermion 
behavior is provided mainly by its superconducting properties, 
like the height of the specific heat jump at the superconducting 
transition or the high value of $H_{c2}(T)$ 
relative to $T_c$\cite{Bauer2002}. PrOs$_{4}$Sb$_{12}$ is cubic with 
$T_h$ point group symmetry \cite{Takegahara2001}, and has a nonmagnetic ground state, which in a single ion scheme 
can be either a $\Gamma_{23}$ doublet 
or a $\Gamma_{1}$ singlet, a question which remains a matter of 
controversy. Presently, most measurements in high field seem to favor a singlet ground 
state \cite{Kohgi2003,Aoki2002,Tayama2003,Rotundu2004}. In any case, whatever the degeneracy of this ground state, the 
first excited state (at around 6K) is low enough to allow for an induced electric quadrupolar moment 
on the Pr$^{3+}$ ions \cite{Maki2003,Harimaprivate}, that could explain the heavy fermion properties of this system by a 
quadrupolar Kondo effect  \cite{Bauer2002}. Thus, while the pairing mechanism of usual HF superconducting 
compounds (U or Ce-based) could come from magnetic fluctuations, the superconducting state of PrOs$_{4}$Sb$_{12}$ 
could be due to quadrupolar fluctuations. Yet at present, this attractive hypothesis is backed by very few 
experimental facts, both as regards the evidence of a quadrupolar Kondo effect in the 
normal phase and as regards the pairing mechanism in the superconducting state.  Even the question of the 
unconventional nature of its superconductivity is still open. Indeed, several types of experiments have already probed the nature of this superconducting state, but with apparently contradictory results. Concerning the gap 
topology, scanning tunneling spectroscopy measurements point to a fully open gap, with some 
anisotropy on the Fermi surface 
\cite{Suderow2004}. Indication of unconventional superconductivity might come from the distribution of 
values of the residual density of states (at zero energy) on 
different parts of the sample surface. 
This could be attributed to a pair-breaking effect of disorder. The 
same conclusion as regards the gap size 
was reached by $\mu SR$ measurements \cite{MacLaughlin2002} and 
$NQR$ measurements \cite{Kotegawa2003}, 
although unconventional superconductivity is suggested in the 
latter case by the absence of a coherence peak below $T_c$ in $1/T_1$. 

This should be contrasted with recent penetration depth 
measurements, that would indicate point nodes of the gap 
\cite{Chia2003}, or the angular dependence of the thermal 
conductivity which suggests an anisotropic 
superconducting gap with a nodal structure \cite{Izawa2003}. This 
latter measurement also suggests multiple 
phases in the temperature ($T$) - field ($H$) plane, which could be 
connected to the double transitions 
observed in zero field  \cite{ Vollmer2003,Maple2002, 
Oeschler2003}. Recent  $\mu SR$ relaxation 
experiments \cite {Aoki2003} detected a broadening of the internal 
field distribution below $T_c$, 
suggesting a multicomponent order parameter or a non unitary odd 
parity state, with a finite magnetic moment. 

In this context, our results bring new insight on the question of the 
parity of the order parameter, and draw a definite picture of the (H,T) phase diagram as deduced 
from specific heat measurements. With reference to the historical case of UPt$_3$, we emphasize that 
the present status of sample quality may explain the discrepancies between the various 
measurements : definite claims on the nature of the superconducting state in PrOs$_4$Sb$_{12}$ are at the 
very best too early, the key point being the sample quality.

\section{Experimental details} 

We present results on single crystals of PrOs$_4$Sb$_{12}$ grown 
by the Sb flux method \cite{Takeda1999,Bauer2001}. These samples 
are aggregates of small single crystals with well developed cubic 
faces. They have a good RRR of about 40 (between room temperature 
and 2K), a superconducting transition  (onset of $C_{p}$ or $\rho$) 
at 1.887~K and, they present a very sharp double 
superconducting transition in the specific heat. 

Two different techniques have been used for the specific heat 
measurements. The first is a quasi-adiabatic method with a 
Au/Fe-Au thermocouple controlled by a superconducting quantum 
interference device (SQUID) in a $^{3}He$ calorimeter. It is well suited to quantitative studies 
in zero field (the addenda are precisely known), and was used on samples with a total mass of 8 mg. 
The second technique was ac calorimetry, used to follow the superconducting transitions under magnetic field 
in order to draw a complete phase diagram of the two superconducting transitions. This ac calorimetry 
uses a strain gauge heater (PtW alloy), a sensitive SiP thermometer (silicon doped with phosphorus close 
to the critical concentration of the metal-insulator transition), and a long gold wire (25$ \mu$m 
diameter) as a heat leak. For the ac method, we choose a frequency of 0.04~Hz and a large integration time of 
350~s. The SiP thermometer is measured with a four lead resistance bridge at 500~Hz, whose analog output 
is sent to the lock-in detection. The heating power was chosen so that the SiP temperature 
oscillations remain smaller than 6~mK in order to avoid broadening of the transitions. Thermometry 
under field was controlled by thermometers located in the (zero field) compensated region of the magnet. 

\section{Specific heat results}

\begin{figure}
\begin{center}
\includegraphics[width =8.5cm]{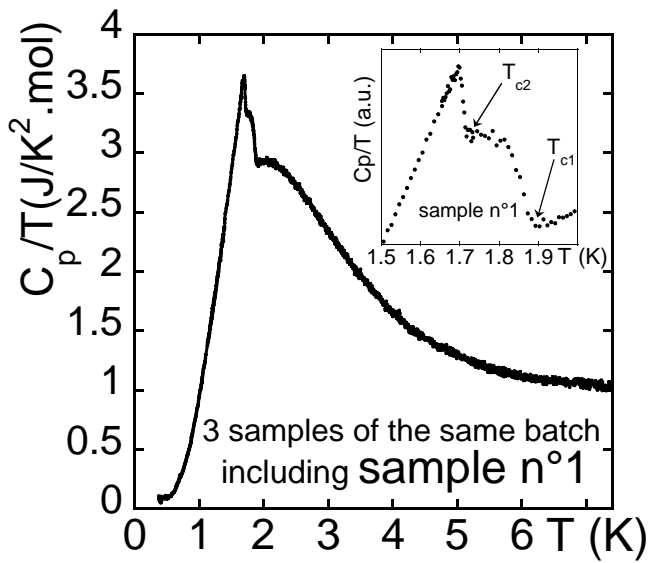}
\caption{Specific heat of samples of the same batch including 
sample $n^\circ1$ as $\displaystyle \frac{C_{p}}{T}$ versus T at 
zero field measured with a quasi-adiabatic method. The inset is a 
zoom on the double superconducting transition of sample $n^\circ1$ 
measured with an ac method: $T_{c1}=1.887~$K and $T_{c2}=1.716~$K.}
\label{Cpnumber1}
\end{center}
\end{figure}

Figure \ref{Cpnumber1} displays the specific heat $C_p(0,T)/T$ of 3 
samples of the same batch measured together, of total mass 8 mg. 
The inset of fig. \ref{Cpnumber1} is the ac specific heat $C_{p}(0,T)$ of one of these three samples 
(thereafter called sample $n^\circ 1$). As it has been previously observed \cite{Bauer2002}, PrOs$_{4}$Sb$_{12}$ 
shows a Schottky anomaly with a maximum in $C_{p}/T$ near $T = 2.2$K. Absolute values of $C_{p}/T$ are 
the highest ever reported: at $T=1.7$K, ${C_p}/{T}$ = $3.65$J/ K$^2$.mol and at $T=2$K, 
${C_p}/{T}$=$2.9$ J/mol.K$^2$. Sample $n^\circ$1 has a well defined double transition: to our knowledge, 
it is the sharpest ever reported in the literature, although we are aware of similar (yet unpublished) 
results by Y. Aoki \cite{Aokiprivate} on samples grown in the same group. The width of the two transitions was estimated 
to be 16mK and 58mK, with respectively $T_{c2} =1.716$K  and $T_{c1}=1.887$K ( with the junction 
criterion). 

\begin{figure}
\begin{center}
%\scalebox{.5}
\includegraphics[width = 8.5cm]{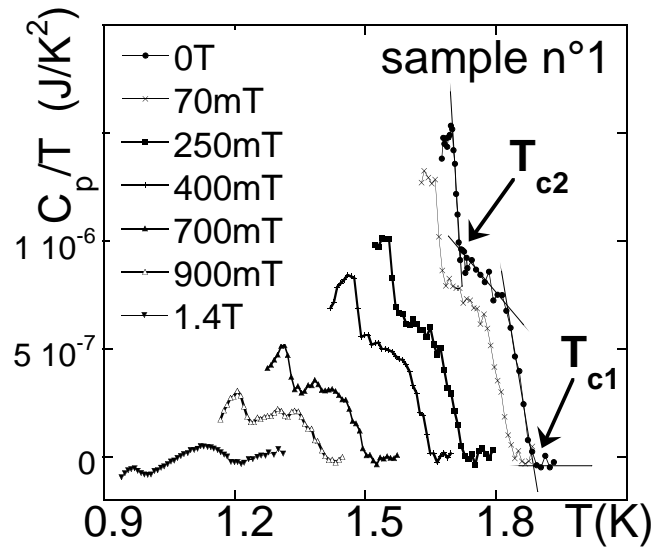}
\caption{ Temperature sweeps of the ac specific 
heat of sample $n^\circ1$ at several fields below 1.4~T. The normal 
state was subtracted. The arrows indicate the double transition $T_{c1}$ and $T_{c2}$.}
\label{rampeT}
\end{center}
\end{figure}

\begin{figure}
\begin{center}
%\scalebox{.5}
\includegraphics[width =8.5cm]{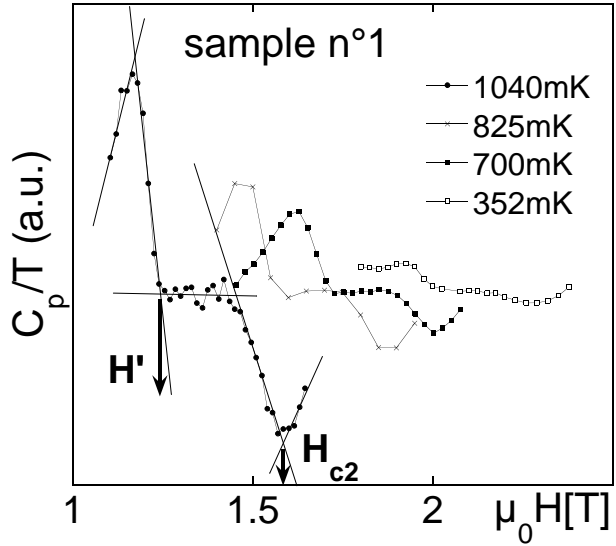}
\caption{Field sweeps of the ac specific 
heat of sample $n^\circ1$ at several temperatures. An arbitrary 
line between the two transitions was subtracted. 
We follow the two transitions ($H_{c2}$ and 
$H^{'}$) down to 350~mK.}
\label{rampeH}
\end{center}
\end{figure}

\begin{figure}
\begin{center}
% \scalebox{.5}
\includegraphics[width =8.5cm]{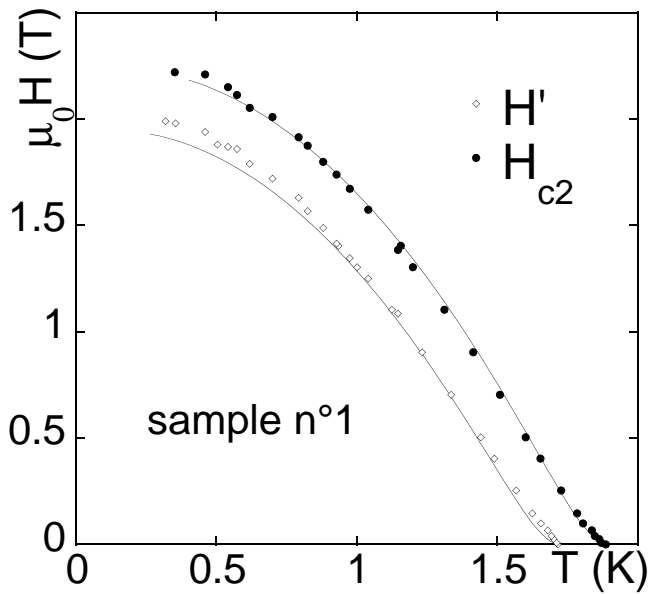}
\caption{$H-T$ superconducting phase diagram of 
PrOs$_{4}$Sb$_{12}$ determined by specific heat measurements on 
sample $n^\circ1$. The field dependence of $T_{c1}$ and $T_{c2}$ are completely similar. The lines are fits by a two-band 
model of the upper critical field (Section~\protect{\ref{sec:model}}).
%$500$mJ/K$^2$.mol  for band 1, 
%$20$mJ/K$^2$.mol for band 2 
Only $T_{c}$ has been changed from $H_{c2}$ to $H^{'}$.}
\label{diagphase}
\end{center}
\end{figure}

Shown in fig. \ref{rampeT} and fig. \ref{rampeH} are the ac specific heat measurements 
$C_{p}(H,T)$ at, respectively, constant magnetic field ($T_{c1}\geq 1.16~$K) and constant temperature ($T_{c1}\leq 
1.15~$K) of the same sample $n^\circ1$. The normal state specific heat, or an arbitrary line between the two 
transitions, has been subtracted for the temperature or field sweeps respectively. 
Even under field, the transitions remain very sharp, so that we were able to detect them down to 350 mK and to draw a 
precise phase diagram (Fig. \ref{diagphase}). 
The width of the two transitions does not exceed $\Delta H_{c2}= 80~$mT and $\Delta H^{'} = 50~$mT at 500~mK.

The phase diagram has the same features as reported by Tayama et al. 
\cite{Tayama2003} from magnetization measurements. The advantage 
of specific heat is to give an unambiguous signature of a bulk 
phase transition, that cannot be confused with other physical 
phenomena like peak-effect. The two transition lines remain 
almost parallel and we will see that they can be deduced from each other simply by scaling $T_{c}$.

\section{Sample characterization}

Three pieces of sample $n^\circ1$ have been used for further 
characterizations, called 1a, 1b and 1c. As well as the specific heat 
of sample $n^\circ1a$ (2~mg), we have measured the resistivity of samples 
$n^\circ1b$ (0.2~mg) and $n^\circ1c$, and the ac susceptibility 
and dc magnetization of samples $n^\circ1a$ and $n^\circ1b$. 

Concerning the specific heat, the high absolute value of $C_p$ as 
well as the large height of the two superconducting jumps ($\Delta 
(C_{p}/T) = 350~$mJ/mol.K$^{2}$ at $T_{c1}$ and $\Delta 
(C_{p}/T)$=$300~$mJ/mol.K$^{2}$ at $T_{c2}$) must be linked to the high quality of these samples 
(absence of Sb-flux and/or good stoichiometry). Moreover, the 
heights of the two steps of sample $n^\circ1a$ are quantitatively similar to those of the 
entire batch (7.5~mg) as Vollmer has already pointed out \cite{Vollmer2003}. 

\begin{figure}
\begin{center}
 %\scalebox{.5}
\includegraphics[width =8.5cm]{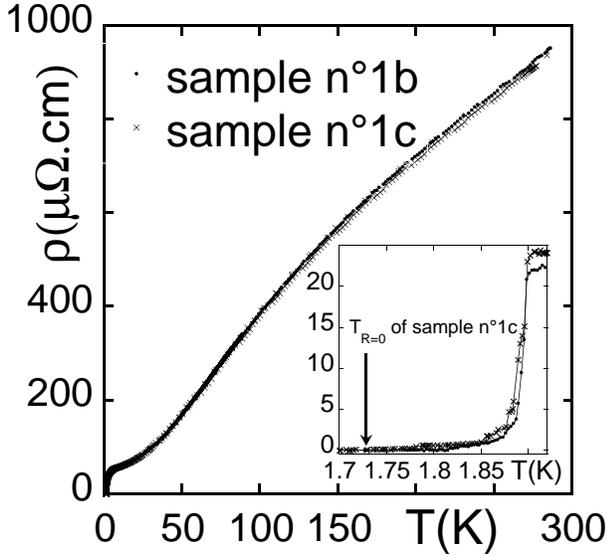}
\caption{ Resistivity of samples $n^\circ1b$ and $n^\circ1c$ 
normalized to the slope at high temperature of sample $n^\circ1c$. The inset 
is a zoom on the superconducting transition. The resistivity of sample $n^\circ1c$ is zero only at $T_{c2}$.}
\label{rho}
\end{center}
\end{figure}

Like in previous work \cite{Bauer2002, Ho2003}, we have noticed that the resistivity at 
300K is very sample dependent ( from 200 to 900 $\mu\Omega$.cm). On all 
samples, the value of $\rho$ at 300~K seems to scale with the 
slope at high temperature ($T \geq 200$K), i.e. the phononic part 
of the resistivity, as if the discrepancies were due to an 
error on the geometric factor. This error could be explained by 
the presence of microcracks in the samples. 
We have taken this problem into account by normalizing all data to 
the slope $d\rho/dT$ at high temperature ($T \geq 200$K) of sample 
$n^\circ$1c, chosen arbitrarily. 
For samples $n^\circ1b$ and $n^\circ1c$ respectively (Fig. \ref{rho}), 
the RRR (ratio between 300K and 2K values) are 44 and 38, the 
onset $T_c$ are 1.899~K and 1.893~K (matching the critical 
temperature obtained by specific heat), and the temperatures of 
vanishing resistance ($T_{R=0}$) are 1.815~K and 1.727~K. 
$T_{R=0}$ of sample $n^\circ1c$ is equal to $T_{c2}$ and this 
remains true under magnetic field. So, in sample $n^\circ1c$, 
the resistive superconducting transition is not complete between 
$T_{c1}$ and $T_{c2}$.

\begin{figure}
\begin{center}
 %\scalebox{.5}
\includegraphics[width =8.5cm]{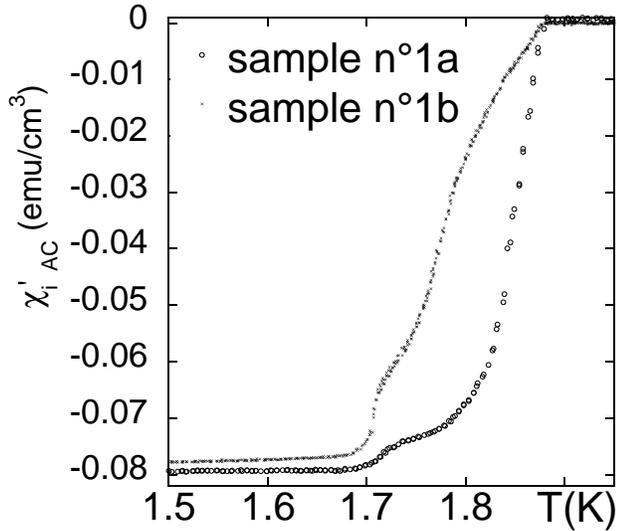}
\caption{Real part $\chi'$ of the AC susceptibility of samples 
$n^\circ1a$ and $n^\circ1b$ measured with an AC magnetic field of 
$0.287$Oe at $2.11$Hz. Like in the results of resistivity 
measurement, the superconducting transition is not complete at 
$T_{c1}$ and the two transitions are visible.}
\label{Xi'}
\end{center}
\end{figure}

\begin{figure}
\begin{center}
 %\scalebox{.5}
\includegraphics[width =8.5cm]{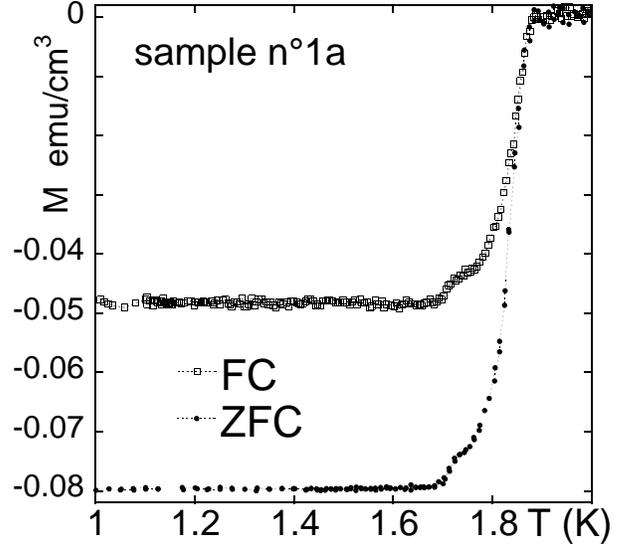}
\caption{DC magnetization of sample $n^\circ1a$ at $H_{DC}=1$Oe 
with zero field cooled (ZFC) and field cooled (FC) sweeps. The 
Meissner effect is $44\%$ for this sample and 
$55\%$ for sample $n^\circ1b$ (not shown). The superconductivity is bulk.}
\label{M-ZFC-FC}
\end{center}
\end{figure}

Figure \ref{Xi'} shows  the superconducting transition for samples 
$n^\circ1a$ and $n^\circ1b$ by ac-susceptibility ($H_{ac}=0.287$Oe), corrected for the 
demagnetization field. 
The onset temperature is the same for the two samples ($1.88$K). 
The transition is complete only at around 1.7K and two transitions 
are visible. The field cooled dc magnetization of samples 
$n^\circ1a$ and $n^\circ1b$ ($H_{dc}=1$Oe), shown in fig.\ref{M-ZFC-FC}, gives a Meissner effect of, respectively, $44\%$ 
and $55\%$, indicating (like specific heat) that the superconductivity is bulk. The two transitions are also visible.

%Thus, the results of resistivity and susceptibility measurements 
%of the superconducting transition can't exclude that the second 
%transition might be due to a part of the sample with a smaller $T_c$. 
%In any case, despite the sharp specific heat jumps at $T_{c1}$ and 
%$T_{c2}$, the superconductivity in sample $n^\circ1$ is still not 
%homogeneous.

\section{Discussion}

\subsection{Double superconducting transition}

Let us first discuss the nature (intrinsic or not) of the double 
transition observed in our specific heat measurements. The 
remarkable fact, compared to previous reports \cite{Vollmer2003,Maple2002}, 
is the progress on the sharpness of both transitions. 
If for previous reports, a simple distribution of $T_c$ due to a 
distribution of strain in the sample could have explained the transition width, it is not 
the case anymore for the sharp features observed on these new samples. Also the explanation recently proposed 
\cite{Broun2004} that the lower transition at $T_{c2}$ would be 
induced by Josephson 
coupling of one sheet of the Fermi surface to another with 
transition temperature 
$T_{c1}$ is excluded, owing to the sharpness of both transitions and 
particularly of 
that at $T_{c2}$: see the broadening calculated by the authors 
\cite{Broun2004} for 
the Josephson induced transition. Nevertheless, some of our 
results still cast 
serious doubts on the intrinsic nature of the double transition. 
Indeed, the 
susceptibility also shows a "double" transition, and resistivity 
becomes zero 
only at $T_{c2}$. If the first transition at $T_{c1}$ was 
bulk-homogeneous, 
the resistivity should immediately sink to zero below  $T_{c1}$, 
and the susceptibility ($\chi$) should also show perfect diamagnetism far 
before $T_{c2}$ : the sample diameter is at least 1000 times larger than the 
penetration depth ($ \lambda$) so that the temperature dependence of $ \lambda$ cannot explain 
such a transition width of 
$\chi$ (contrary to the statements of E.E.M. Chia \cite{Chia2003}). So both 
resistivity and susceptibility 
are indicative of remaining sample inhomogeneities.

Nevertheless, in our opinion, even the comparison of the various 
characterizations of the superconducting transition by resistivity, susceptibility and 
specific heat on the same sample does not allow for a definite conclusion. 

Two historical cases are worth remembering. URu$_2$Si$_2$ showed a double transition in 
the specific heat in some samples, with inhomogeneous features detected in the 
susceptibility \cite{Ramirez1991}. In that case, the authors of reference \cite{Ramirez1991} could 
clearly show that it was not intrinsic (maybe arising from internal strain ?) 
because different macroscopic parts of the same sample showed one or the other 
transition. In PrOs$_4$Sb$_{12}$, the specific heat results are 
reproducible among various samples of the same batch (see samples $n^\circ1$ and 
$n^\circ1a$ of the present work), and such an easy test does not work.

The second case is of course UPt$_3$: it is now well established that the two transitions 
observed in zero field are intrinsic and correspond to order parameters of different 
symmetries. But the first results on samples that were not homogeneous enough 
showed exactly the same behaviour as the present  one in PrOs$_4$Sb$_{12}$: two 
features in the susceptibility and a very broad (covering both transitions) 
resistive transition \cite{Sulpice1986}. It was not until the sample quality improved 
significantly that resistivity and susceptibility 
transitions matched the higher one \cite{Hasselbach1989}. The puzzling result for PrOs$_4$Sb$_{12}$, 
compared to UPt$_3$, is that despite the sharpness of the 
specific heat transitions, resistivity and susceptibility reveal 
inhomogeneities, which means that this new compound probably has unusual 
metallurgical specificities.

Continuing the parallel with UPt$_3$, basic measurements rapidly 
supported the intrinsic origin of the two transitions: they were probing the respective field 
and pressure dependence of both transitions. Indeed, a complete (H-T) phase 
diagram was rapidly drawn, showing that in UPt$_3$ \cite{Adenwala1990}, like in 
$U_{1-x}Th_{x}Be_{13}$\cite{Jin1996}, the two transitions observed in zero field eventually merged under 
magnetic field, due to a substantial difference in $dT_c/dH$. It is even more true 
for the pressure dependence of $T_{c1,2}$, as the thermal dilation has jumps of 
opposite sign at the two transitions, indicating opposite variations of $T_{c1,2}$ 
under pressure (Ehrenfest relations) \cite{Hasselbach1990}. So in this compound, 
the two transitions 
could be rapidly associated with a change of the symmetry of the 
order parameter indicating the unconventional nature of the superconducting 
state. 

We are not so lucky in the case of PrOs$_4$Sb$_{12}$: indeed, the 
field dependence of $T_{c2}$ seems completely similar to that of 
$T_{c1}$  (fig. \ref{diagphase}), a claim that will be made 
quantitative below. It is the same situation as in 
URu$_2$Si$_2$ \cite{Ramirez1991}, and nothing in favor of an 
intrinsic nature of the double transition can be deduced from this phase diagram. 
Another phase diagram has already been established by transport measurements, with a line 
$H^*(T)$ separating regions of twofold and fourfold symmetry  in the angular dependence 
of thermal conductivity under magnetic field \cite{Izawa2003}. It may seem likely 
\cite{Izawa2003, JunGoryo2003} that this line would merge with the double transition in zero 
field. From the line $H^{'}(T)$ drawn from our specific heat measurements ($T_{c2}(H)$, fig. \ref{diagphase}), we can 
onclude that this is not the case : $H^{'}(T)$  does not match the line $H^*(T)$ drawn 
in reference \cite{Izawa2003}, unless there is an unlikely strong sample dependence of these lines.

Comparison of the pressure dependence of  $T_{c1}$ and  $T_{c2}$ 
seems more promising: contrary to case of UPt$_{3}$ \cite{Hasselbach1990} the jump of the 
thermal expansion at the two superconducting temperatures does not change sign 
\cite{Oeschler2003}, but from the relative magnitude of these jumps and our specific 
heat peaks, we get a value $dTc_1/dp\approx -0.2$K/GPa, and twice as much for 
$dTc_2/dp$. This supports a different origin for both transitions. The weak point is that 
the thermal expansion measurements were done on two samples mounted on top of each 
other, that were early samples with rather broad specific heat transitions. Thus, the 
question of the intrinsic nature of the double superconducting transition remains open.

\subsection {Upper critical field }

Another quantity which has not been thoroughly discussed is the upper 
critical field $H_{c2}(T)$. In heavy fermion superconductors, $H_{c2}(T)$ has always 
proved to be an interesting quantity, mainly due to the large mass enhancement of the 
quasiparticles. Indeed, the usual orbital limitation is very high in these systems, 
due to the low Fermi velocity, so that the authors of reference \cite{Bauer2002} 
could, from their $H_{c2}(T)$ data, confirm the implication of heavy quasiparticles in 
the Cooper pairs (revealed also by the specific heat jump at $T_c$). They also gave 
an estimate of the heaviest mass: $\approx 50$ m$_0$, where m$_0$ is the free electron 
mass.

But, as the orbital limitation is very high, $H_{c2}(T)$ 
is also controlled by the paramagnetic 
effect. A quantitative fit of $H_{c2}(T)$ easily 
gives the amount of both limitations, except that on this system, 
$H_{c2}(T)$ has an extra feature: a small initial positive curvature close to $T_c$. 
This feature has been systematically found, whatever the samples and the techniques 
used to determine $H_{c2}(T)$ ($\rho$, $\chi$ and $C_{p}$) \cite{Bauer2002, Izawa2003, Vollmer2003, Ho2003}. 
Our data, obtained by $\rho$ on sample $n^\circ1c$ and $C_p$ 
on sample $n^\circ1$, matches all other published 
results. So we can now consider this curvature of H$_{c2}(T)$ as 
intrinsic, and not due to some artifact of transport measurements, or coming from 
inhomogeneity in the sample. Such an intrinsic positive curvature also appears in MgB$_{2}$ 
\cite{budko2002} or in borocarbides like YNi$_{2}$B$_{2}$C and LuNi$_{2}$B$_{2}$C \cite{Shulga1998}. 

\subsubsection {Physical inputs}

We propose an explanation based on different gap amplitudes for the 
different sheets of the Fermi surface of PrOs$_{4}$Sb$_{12}$, which is made 
quantitative through a "two-band" model \cite{Prohammer1987} . Microscopically, STM 
spectroscopy also reveals an anisotropic gap, with zero density of states at low 
energy (fully open gap) but a large smearing of the spectra \cite{Suderow2004}. Recent 
microwave spectroscopy measurements also discuss a two band model \cite{Broun2004}, but 
in order to explain the double transition. We insist that our model has nothing to do 
with the double transition, which clearly involves heavy quasiparticles both at 
$T_{c1}$ and at $T_{c2}$ (see the size of the two specific heat jumps): our aim 
is a quantitative understanding of $H_{c2}(T)$, based on the normal state 
properties of PrOs$_{4}$Sb$_{12}$, as in MgB$_2$ or borocarbides where no 
double transition has ever been observed. The physical input of a multi-band 
model for $H_{c2}(T)$ is to introduce different Fermi velocities and 
different inter and intra band couplings.  As a result, $T_{c}$ is 
always larger than for any of the individual bands \cite{Mazin2003}. The 
slope of $H_{c2}$ at $T_{c}$ is larger for slower Fermi velocity 
(heavier) bands. Positive curvature of $H_{c2}(T)$ is easily obtained if 
the strongest coupling is in the heaviest bands (large $H_{c2}$), with a 
slight $T_{c}$ increase due to the inter band coupling to the lightest 
bands (small initial slope) \cite{Shulga1998}.

\subsubsection{\label{sec:model}The two-band model:}

There are at least three sheets for the Fermi surface of 
PrOs$_4$Sb$_{12}$, but in the absence of a precise knowledge of 
the pairing interaction, a full model would be unrealistic, having 
an irrelevant number of free parameters. 
A two band model is enough to capture the physics of anisotropic 
pairing, although only the correspondence with band calculations 
becomes looser. In our model, band 2 would correspond to the lightest ($\beta$) band detected by 
de-Haas van Alphen measurements, and band 1 would be a heavy band having 
most of the density of states.  Indeed, the de Haas-van Alphen 
experiments on PrOs$_{4}$Sb$_{12}$ \cite{Sugawara2003,Sugawara2002} reveal the 
presence of light quasiparticles (band  $\beta$) and heavier particles 
(band $\gamma$). The heaviest quasiparticles are at present only seen by 
thermodynamic measurements ($C_{p}$ or $H_{c2}$). Anisotropic coupling 
between the quasiparticles is considered in the framework of an Eliashberg 
strong coupling two-band model \cite{Prohammer1987} in the 
clean limit, with an Einstein phonon spectrum (characteristic "Debye" frequency $\Omega$). 
Let us point out that the results do not depend on (and a 
fortiori do not probe) the pairing mechanism, which is likely to be much more 
exotic than the usual electron-phonon mechanism. Compared to a single 
band calculation, there is now a matrix of strong coupling parameters 
$\lambda_{i,j}$ describing the diffusion of electrons of band i to band j by the excitations responsible for the pairing. 

What matters for $H_{c2}(T)$ is the relative weight of the 
$\lambda_{i,j}$, not their absolute value: we consider $T_{c}$ or the renormalized 
Fermi velocities as experimental inputs.  $\lambda_{i,j}$ depends both on 
the interaction matrix elements between bands $i$ and $j$, and on the 
final density of states of band $j$ \cite{Mazin2003}. In MgB$_{2}$, it is claimed that 
electron-phonon coupling is largest within the $\sigma$ bands. 
Here, knowing nothing about the pairing mechanism, we assume constant 
inter and intra band coupling, so that the relative weight of the 
$\lambda_{i,j}$ is only governed by the density of states of band $j$. This density 
of states is itself proportional to the contribution of that band to the specific 
heat Sommerfeld coefficient: $500$mJ/K$^2$.mol \cite{Bauer2002} for band 1, 
$20$mJ/K$^2$.mol for band 2 \cite{Sugawara2003,Sugawara2002}. The Fermi 
velocity of band 1 (not observed by de Haas-van Alphen measurements) is the main 
adjustable parameter of the fit : we find $v_{F1}=0.0153~10^{6}$m/s$^{-1}$, in 
agreement with \cite{Bauer2002} where the Fermi velocity has been inferred 
from the slope of $H_{c2}(T)$ at $T_{c}$ ignoring the initial positive curvature. 
All other coefficients are either arbitrary ($\lambda_{1,1}$=1) 
(in agreement with the strong coupling superconductivity 
concluded in \cite{Kotegawa2003}), conventional values (gyromagnetic ratio 
for the paramagnetic limitation $g=2$, Coulomb repulsion parameter 
$\mu^{*}_{i,j}=0.1\delta_{i,j}$), or fixed by experimental data 
($T_{c}=1.887~$K$ \Longrightarrow \Omega=21.7$K, 
$v_{F2}=0.116.10^{6}$m.s$^{-1}$ from the de Haas-van Alphen data on 
the $\beta$ band). The model fits well the experimental data (cf. 
fig. \ref{diagphase}), including the small positive curvature. 
Before discussing the interpretation of the fit as regards the 
values of the parameters and the parity of the superconducting 
order parameter, let us note that we can fit the $H^{'}(T)$ line 
(fig. \ref{diagphase}) with the same parameters as for $H_{c2}$ except $\Omega$, 
adjusted to give $T_{c}=1.716$K. There is a good agreement with all data except at very low temperature or near 
$T_{c}$ where the curvature is reduced compared to $H_{c2}(T)$. 
However, these deviations are weak, and this is why we claim that 
$H^{'}(T)$ has the same behavior than $H_{c2}(T)$, which does not 
help to identify the second transition with a symmetry change of 
the superconducting order parameter.

\subsubsection {Interpretation}

\begin{figure}
\begin{center}
 %\scalebox{.5}
\includegraphics*[width =8.5cm]{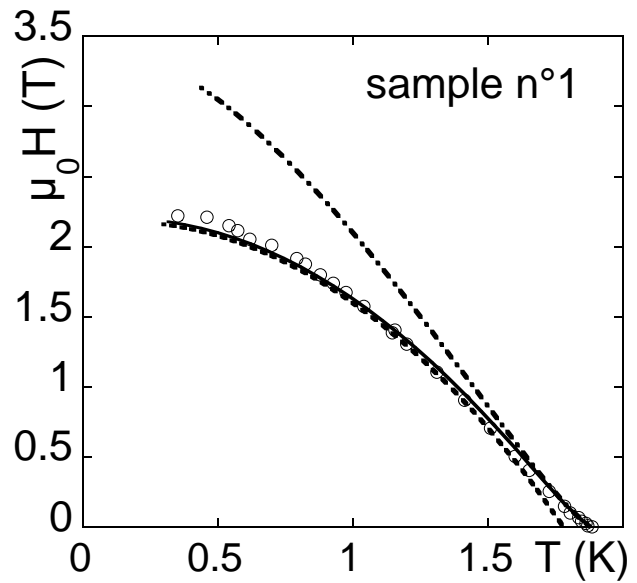}
\caption{Open circles show the data of $H_{c2}(T)$ by specific heat 
measurement on PrOs$_{4}$Sb$_{12}$ (sample  $n^\circ1$) . The lines show fits with 
a two-band model (solid line), with a single-band model (dashed line), without the paramagnetic 
limit (g=0) (dotted line). It shows that the increase of $T_{c}$ due 
to the coupling with light qp band is rapidly suppressed in weak 
magnetic fields. $H_{c2}$ is also clearly Pauli limited, supporting 
a singlet superconducting state. The parameters for the solid line fit are: $g=2$, $\mu^{*}_{i,j}=0.1\delta_{i,j}$, 
$v_{F1}=0.0153~10^{6}$m/s$^{-1}$ (heavy band), $v_{F2}=0.116.10^{6}$m.s$^{-1}$ (lightest band, from de Haas-van Alphen oscillations \protect{\cite{Sugawara2002}}), $T_{c}=1.887~$K$ \Longrightarrow \Omega=21.7$K.}
\label{diagrphase1band}
\end{center}
\end{figure}

Shown also in fig. \ref{diagrphase1band} are the calculations of 
$H_{c2}(T)$ for a single band model: $v_{F1}$, the characteristic frequency 
$\Omega$ and $\lambda_{1,1}$ have the same values as before, but all 
other $\lambda_{i,j}$ coefficients have been turned down 
to zero, eliminating the effects of the light electron band. 
$T_{c}$ is then reduced (down to 1.78~K) and the positive curvature 
disappears.  We also observe that the fit of $H_{c2}$ is basically 
unchanged above 1~T, meaning that low fields suppress the 
superconductivity due to the light electron band restoring a "single 
band" superconducting state. This is the same effect observed 
more directly in MgB$_{2}$ with specific heat measurements under 
magnetic field : the smaller gap rapidly vanishes, leading to a 
finite density of states at the Fermi level under magnetic fields 
due to the $\pi$ band \cite{Bouquet2002}. This suppression of 
the light quasiparticle superconductivity would have here an effect 
on specific heat too small to be observed (contribution of 
the light quasiparticles to the specific heat of order $4\%$ of the 
Sommerfeld coefficient, itself buried in the large Schottky anomaly). 
But it may have much larger effects on transport. 
Let us note that the clean limit is a posteriori justified: 
from $v_{F1}$, we find a coherence length $\xi_{0}\sim110\AA$, 
whereas from a residual resistivity $\rho_{0}$ and specific heat 
coefficient $\gamma \sim0.5$J/K$^{2}$.mol$\sim2.10^{3}$J/K$^{2}$.m$^{3}$, 
we get $v_{F1}l\sim\displaystyle\frac{3L_{0}}{\rho\gamma}\sim2.10^{-3}$m$^{2}$/s yielding a mean free path 
$l \sim 1300\AA>10\xi_{0}$. 

More interestingly, the quantitative fit of $H_{c2}$ also allows
 a discussion of the parity of the order parameter in 
PrOs$_{4}$Sb$_{12}$. Indeed, like other heavy fermion superconductors, despite the 
low-T$_{c}$ value, $H_{c2}$ can be sensitive to the Pauli limit in case of 
singlet pairing. The fit in fig. \ref{diagrphase1band} includes this 
paramagnetic limitation, with the conventional free electron value for $g$ 
($g=2$). Also shown in fig. \ref{diagrphase1band} is the calculation of 
$H_{c2}(T)$ with the same parameters but for $g=0$, i.e. without any paramagnetic 
limit. The strong deviations observed demonstrate that the paramagnetic 
effect controls $H_{c2}$ at low T in PrOs$_{4}$Sb$_{12}$. Quantitatively, 
the paramagnetic effect depends on the coupling strength. Yet, we 
choose arbitrarily $\lambda=1$. A rather strong coupling regime is 
supported by $NQR$ experiments \cite{Kotegawa2003}. Even for a weak 
coupling picture ($\lambda=0.6$), the fit yields $g=1.55$. In both cases, 
the paramagnetic limit remains important, supporting a singlet 
nature of the superconductivity, contrary to that had been suggested in 
\cite{Anders2002, Miyake2003, Ichioka2003} and in agreement with 
the supposition in \cite{Maki2003, JunGoryo2003}. This result should be quite robust, independent of 
the two band model. It could 
be invalidated if the mass renormalization mechanism was field 
dependent, which could be an explanation for the difference in the 
large specific heat Sommerfeld coefficient obtained in low field, and the 
de Haas-van Alphen measurements performed at high field 
\cite{Harimaprivate}. In such a case, the "saturation" of $H_{c2}(T)$ at low 
temperature could arise from a reinforcement of the orbital 
limitation alone.

\section{conclusion}

To conclude, we have drawn a very precise superconducting phase 
diagram of PrOs$_4$Sb$_{12}$ down to 350~mK, by a specific heat 
measurement. We have yet no clear evidence of the unconventional 
nature of the superconducting order parameter from this phase 
diagram. The superconducting phase diagram with the symmetry 
change of the order parameter drawn by K. Izawa et al 
\cite{Izawa2003} from transport measurements does not seem related 
to the double transition observed with specific heat measurements. 
Despite the high quality of the sample, we cannot completely 
exclude that there are still two parts with different $T_{c}$ in 
our sample, as $H^{'}(T)$ is just scaled from $H_{c2}$ with 
respect to $T_c$. The puzzling result is that despite sharp 
specific heat transitions, inhomogeneities are still present in 
the samples. This calls for caution in the claim of various types 
of nodes of the gap by different sophisticated techniques: the 
most urgent task is to understand the problem of sample quality. 
Contrastingly, the upper critical field is very reproducible, 
independent of samples and types of measurements. It has been 
analyzed with a strong coupling two-band model taking account of 
the spread in the effective masses of the quasi-particles and of 
the pairing strength as suggested also by STM spectroscopy 
measurements \cite{Suderow2004}. The strong influence of the 
paramagnetic limit on $H_{c2}$ is the first experimental argument 
for a singlet superconducting order parameter.

\acknowledgments We acknowledge many fruitfull discussions with K.
Izawa, H. Harima, V. Mineev, H. Suderow, J-L. Tholence,  P.C.
Canfield and G. Lapertot. 
This research was supported by the Grant-in Aid for Scientific Research on the Priority Area 
"Skutterudites" from MEXT in Japan.


\begin{thebibliography}{99}




\bibitem{Bauer2002}
E.D. Bauer, N.A. Frederick, P.-C. Ho, V.S. Zapf, and M.B. Maple, Phys. Rev. B \textbf{65}, 100506(R) (2002).

\bibitem{Takegahara2001}
K. Takegahara, H. Harima and A. Yanase, J. Phys. Soc. Jpn.
\textbf{70}, 1190 (2001).

\bibitem{Kohgi2003}
M. Kohgi, K. Iwasa, M. Nakajima, N. Metoki, S. Araki, 	N. Bernhoeft, J.-M. Mignot, A. Gukasov, H. Sato, Y. Aoki and H. Sugawara, J. Phys. Soc. Jpn. \textbf{72}, 1002 (2003).

\bibitem{Aoki2002}
Y. Aoki, T. Namiki, S. Ohsaki, S.R. Saha, H. Sugawara, and H. Sato, J. Phys. Soc. Jpn. \textbf{71}, 2098 (2002).

\bibitem{Tayama2003}
T. Tayama, T. Sakakibara, H. Sugawara, Y. Aoki, and H. Sato, J. Phys. Soc. Jpn.  \textbf{72}, 1516 (2003).

\bibitem {Rotundu2004}
C.R. Rotundu, H. Tsujii, Y. Takano, B. Andraka, H. Sugawara, Y. Aoki, and H. Sato, Phys. Rev. Letters \textbf{92}, 037203
(2004).

\bibitem{Maki2003}
K. Maki, H. Won, P. Thalmeier, Q. Yuan, K. Izawa and Y. Matsuda,
Europhys. Letters. \textbf{64}, 496, (2003).

\bibitem{Harimaprivate} H. Harima (private communication).

\bibitem{Suderow2004}
H. Suderow, S. Vieira, J.D. Strand, S. Bud'ko and P.C. Canfield, Phys. Rev. B \textbf{69}, 060504 (2004).

\bibitem{MacLaughlin2002}
D.E. MacLaughlin, J.E. Sonier, R.H. Heffner, O.O. Bernal, B.L. Young, M.S. Rose, G.D. Morris, E.D. Bauer, T.D. Do and M.B. Maple, Phys. Rev. Letters \textbf{89}, 157001 (2002).

\bibitem{Kotegawa2003}
H. Kotegawa, M. Yogi, Y. Imamura, Y. Kawasaki, G.-q. Zheng, Y.
Kitaoka, S. Ohsaki, H. Sugawara, Y. Aoki, and H. Sato, Phys. Rev.
Letters \textbf{90}, 027001 (2003).

\bibitem {Chia2003}
E.E.M. Chia, M.B. Salamon, H. Sugawara, and H. Sato, Phys. Rev.
Letters \textbf{91}, 247003 (2003).

\bibitem{Izawa2003}
K. Izawa, Y. Nakajima, J. Goryo, Y. Matsuda, S. Osaki, H. Sugawara, H. Sato, P. Thalmeier, and K. Maki, Phys. Rev. Letters
\textbf{90}, 117001 (2003).

\bibitem{Vollmer2003}
R. Vollmer, A. Fai\ss t, C. Pfleiderer, H. v.
L\"{o}hneysen, E.D. Bauer, P.-C. Ho, V. Zapf and M.B. Maple, Phys. Rev. Lett.
\textbf{90}, 057001 (2003).

\bibitem {Maple2002}
M.B. Maple, P.-C. Ho, V.S. Zapf, N.A. Frederick, E.D. Bauer, W.M. Yuhasz, F.M. Woodward, and J.W. Lynn, J. Phys. Soc.
Jpn. \textbf{71} (2002), Suppl. 23.

\bibitem {Oeschler2003}
N. Oeschler, P. Gegenwart, F. Steglich, N.A. Frederick, E.D. Bauer, and M.B. Maple, Acta Physica Polonica B \textbf{34}, 959 (2003).

\bibitem{Aoki2003}
Y. Aoki, A. Tsuchiya, T. Kanayama, S.R. Saha, H. Sugawara,
H. Sato, W. Higemoto, A. Koda, K. Ohishi, K. Nishiyama, and R. Kadono, Phys. Rev. Letters \textbf{91}, 067003 (2003).


\bibitem{Takeda1999}
N. Takeda and M. Ishikawa, Physica B \textbf{259-261}, 92 (1999).

\bibitem{Bauer2001}
E.D. Bauer, A. \'{S}lebarski, E.J. Freeman, C. Sirvent, and M.B.
Maple: J. Phys: Condens Matter \textbf{13}, 4495-4503 (2001).

\bibitem{Aokiprivate} Y. Aoki (private communication).

\bibitem {Ho2003}
P.-C. Ho, N.A. Frederick, V.S. Zapf, E.D. Bauer, T.D. Do, M.B. Maple, A.D. Christianson, and A.H. Lacerda, Phys. Rev. B
\textbf{67}, 180508 (2003).

\bibitem{Broun2004}
D.M. Broun, P.J. Turner, G.K. Mullins, D.E. Sheehy, X.G. Zheng, S.K. Kim, N.A. Frederick, M.B. Maple, W.N. Hardy, and D.A. Bonn, cond. mat. 0310613

\bibitem{Ramirez1991}
A.P. Ramirez, T. Siegrist, T.T.M. Palstra, J.D. Garrett, E. Bruck, A.A. Menovsky and J.A. Mydosh, Phys. Rev. B \textbf{44}, 5392
(1991).

\bibitem{Sulpice1986}
A. Sulpice, P. Gandit, J. Chaussy, J. Flouquet, D. Jaccard, P. Lejay and J.L. Tholence, J. Low Temp. Phys \textbf{62}, 39 (1986).

\bibitem{Hasselbach1989}
K. Hasselbach, L. Taillefer and J. Flouquet, Phys. Rev. Letters
\textbf{63}, 93 (1989).

\bibitem{Adenwala1990}
G. Bruls, D. Weber, B. Wolf, P. Thalmeier, B. L\"{u}thi, A. de Visser and A. Menovsky, Phys.  Rev.  Letters
{\bf 65}, 2294 (1990);\\S. Adenwalla, S.W. Lin, Q.Z. Ran, Z. Zhao, J.B. Ketterson, J.A. Sauls, L. Taillefer, D.G. Hinks, M. Levy and B.K. Sarma, Phys.  Rev.  Letters  {\bf 65}, 2298 (1990).

\bibitem {Jin1996}
D.S. Jin, S.A. Carter, T.F. Rosenbaum, J.S. Kim, and G.R. Stewart,
Phys. Rev. B \textbf{53}, 8549 (1996).

\bibitem{Hasselbach1990}
K. Hasselbach, A. Lacerda, K. Behnia, L. Taillefer, and J. Flouquet,
J. of Low Temp. Phys. \textbf{81}, 299 (1990).

\bibitem{JunGoryo2003}
Jun Goryo, Phys. Rev. B \textbf{67}, 184511 (2003).

\bibitem{budko2002}
S.L. Budko and P.C. Canfield, Phys. Rev. B \textbf{65}, 212501
(2002).

\bibitem{Shulga1998}
S.V. Shulga, S.-L. Drechsler, G. Fuchs, K.-H. M\"{u}ller, K.
Winzer, M. Heinecke, and K. Krug, Phys. Rev. Letters \textbf{80}, 1730
(1998).

\bibitem{Prohammer1987}
M. Prohammer and E. Schachinger, Phys. Rev. B \textbf{36}, 8353 (1987).

\bibitem {Mazin2003}
I.I. Mazin and V.P. Antropov, Physica C \textbf{385}, 49 (2003).

\bibitem {Sugawara2003}
H. Sugawara, S. Osaki, S.R. Saha, Y. Aoki, and H. Sato, Acta Physica
Polonica B \textbf{34}, 1125 (2003).

\bibitem {Sugawara2002}
H. Sugawara, S. Osaki, S.R. Saha, Y. Aoki, H. Sato, Y. Inada, H. Shishido, R. Settai, Y. Onuki, H. Harima, and K. Oikawa, Phys. Rev. B \textbf{66}, 220504(R) (2002).

\bibitem {Bouquet2002}
F. Bouquet, Y. Wang, I. Sheikin, T. Plackowski, A. Junod, S. Lee and S. Tajima, Phys. Rev. Letters \textbf{89}, 257001 (2002).

\bibitem {Anders2002}
F.B. Anders, Eur. Phys. J. B \textbf{28}, 9-28 (2002).


\bibitem{Miyake2003}
K. Miyake, H. Kohno, and H. Harima, J. of Phys. Cond. Matter
\textbf{15}, L275-84 (2003).

\bibitem{Ichioka2003}
M. Ichioka, N. Nakai, and K. Machida, J. of the Phys. Soc. of
Japan \textbf{72}, 1322 (2003).


\end{thebibliography}
\end{document}